\documentclass[aps,prl,twocolumn,showpacs,superscriptaddress,groupedaddress]{revtex4}  
\usepackage{graphicx}  
\usepackage{dcolumn}   
\usepackage{bm}        
\usepackage{epsfig}

\newcommand{\singletop} {\sc singletop}

\begin{document}
\hspace{5.2in} \mbox{FERMILAB-PUB-11/004-E}
\title{
Search for $W{^\prime}\rightarrow tb$ resonances with 
left- and right-handed couplings to fermions

}

\affiliation{Universidad de Buenos Aires, Buenos Aires, Argentina}
\affiliation{LAFEX, Centro Brasileiro de Pesquisas F{\'\i}sicas, Rio de Janeiro, Brazil}
\affiliation{Universidade do Estado do Rio de Janeiro, Rio de Janeiro, Brazil}
\affiliation{Universidade Federal do ABC, Santo Andr\'e, Brazil}
\affiliation{Instituto de F\'{\i}sica Te\'orica, Universidade Estadual Paulista, S\~ao Paulo, Brazil}
\affiliation{Simon Fraser University, Vancouver, British Columbia, and York University, Toronto, Ontario, Canada}
\affiliation{University of Science and Technology of China, Hefei, People's Republic of China}
\affiliation{Universidad de los Andes, Bogot\'{a}, Colombia}
\affiliation{Charles University, Faculty of Mathematics and Physics, Center for Particle Physics, Prague, Czech Republic}
\affiliation{Czech Technical University in Prague, Prague, Czech Republic}
\affiliation{Center for Particle Physics, Institute of Physics, Academy of Sciences of the Czech Republic, Prague, Czech Republic}
\affiliation{Universidad San Francisco de Quito, Quito, Ecuador}
\affiliation{LPC, Universit\'e Blaise Pascal, CNRS/IN2P3, Clermont, France}
\affiliation{LPSC, Universit\'e Joseph Fourier Grenoble 1, CNRS/IN2P3, Institut National Polytechnique de Grenoble, Grenoble, France}
\affiliation{CPPM, Aix-Marseille Universit\'e, CNRS/IN2P3, Marseille, France}
\affiliation{LAL, Universit\'e Paris-Sud, CNRS/IN2P3, Orsay, France}
\affiliation{LPNHE, Universit\'es Paris VI and VII, CNRS/IN2P3, Paris, France}
\affiliation{CEA, Irfu, SPP, Saclay, France}
\affiliation{IPHC, Universit\'e de Strasbourg, CNRS/IN2P3, Strasbourg, France}
\affiliation{IPNL, Universit\'e Lyon 1, CNRS/IN2P3, Villeurbanne, France and Universit\'e de Lyon, Lyon, France}
\affiliation{III. Physikalisches Institut A, RWTH Aachen University, Aachen, Germany}
\affiliation{Physikalisches Institut, Universit{\"a}t Freiburg, Freiburg, Germany}
\affiliation{II. Physikalisches Institut, Georg-August-Universit{\"a}t G\"ottingen, G\"ottingen, Germany}
\affiliation{Institut f{\"u}r Physik, Universit{\"a}t Mainz, Mainz, Germany}
\affiliation{Ludwig-Maximilians-Universit{\"a}t M{\"u}nchen, M{\"u}nchen, Germany}
\affiliation{Fachbereich Physik, Bergische Universit{\"a}t Wuppertal, Wuppertal, Germany}
\affiliation{Panjab University, Chandigarh, India}
\affiliation{Delhi University, Delhi, India}
\affiliation{Tata Institute of Fundamental Research, Mumbai, India}
\affiliation{University College Dublin, Dublin, Ireland}
\affiliation{Korea Detector Laboratory, Korea University, Seoul, Korea}
\affiliation{CINVESTAV, Mexico City, Mexico}
\affiliation{FOM-Institute NIKHEF and University of Amsterdam/NIKHEF, Amsterdam, The Netherlands}
\affiliation{Radboud University Nijmegen/NIKHEF, Nijmegen, The Netherlands}
\affiliation{Joint Institute for Nuclear Research, Dubna, Russia}
\affiliation{Institute for Theoretical and Experimental Physics, Moscow, Russia}
\affiliation{Moscow State University, Moscow, Russia}
\affiliation{Institute for High Energy Physics, Protvino, Russia}
\affiliation{Petersburg Nuclear Physics Institute, St. Petersburg, Russia}
\affiliation{Stockholm University, Stockholm and Uppsala University, Uppsala, Sweden }
\affiliation{Lancaster University, Lancaster LA1 4YB, United Kingdom}
\affiliation{Imperial College London, London SW7 2AZ, United Kingdom}
\affiliation{The University of Manchester, Manchester M13 9PL, United Kingdom}
\affiliation{University of Arizona, Tucson, Arizona 85721, USA}
\affiliation{University of California Riverside, Riverside, California 92521, USA}
\affiliation{Florida State University, Tallahassee, Florida 32306, USA}
\affiliation{Fermi National Accelerator Laboratory, Batavia, Illinois 60510, USA}
\affiliation{University of Illinois at Chicago, Chicago, Illinois 60607, USA}
\affiliation{Northern Illinois University, DeKalb, Illinois 60115, USA}
\affiliation{Northwestern University, Evanston, Illinois 60208, USA}
\affiliation{Indiana University, Bloomington, Indiana 47405, USA}
\affiliation{Purdue University Calumet, Hammond, Indiana 46323, USA}
\affiliation{University of Notre Dame, Notre Dame, Indiana 46556, USA}
\affiliation{Iowa State University, Ames, Iowa 50011, USA}
\affiliation{University of Kansas, Lawrence, Kansas 66045, USA}
\affiliation{Kansas State University, Manhattan, Kansas 66506, USA}
\affiliation{Louisiana Tech University, Ruston, Louisiana 71272, USA}
\affiliation{Boston University, Boston, Massachusetts 02215, USA}
\affiliation{Northeastern University, Boston, Massachusetts 02115, USA}
\affiliation{University of Michigan, Ann Arbor, Michigan 48109, USA}
\affiliation{Michigan State University, East Lansing, Michigan 48824, USA}
\affiliation{University of Mississippi, University, Mississippi 38677, USA}
\affiliation{University of Nebraska, Lincoln, Nebraska 68588, USA}
\affiliation{Rutgers University, Piscataway, New Jersey 08855, USA}
\affiliation{Princeton University, Princeton, New Jersey 08544, USA}
\affiliation{State University of New York, Buffalo, New York 14260, USA}
\affiliation{Columbia University, New York, New York 10027, USA}
\affiliation{University of Rochester, Rochester, New York 14627, USA}
\affiliation{State University of New York, Stony Brook, New York 11794, USA}
\affiliation{Brookhaven National Laboratory, Upton, New York 11973, USA}
\affiliation{Langston University, Langston, Oklahoma 73050, USA}
\affiliation{University of Oklahoma, Norman, Oklahoma 73019, USA}
\affiliation{Oklahoma State University, Stillwater, Oklahoma 74078, USA}
\affiliation{Brown University, Providence, Rhode Island 02912, USA}
\affiliation{University of Texas, Arlington, Texas 76019, USA}
\affiliation{Southern Methodist University, Dallas, Texas 75275, USA}
\affiliation{Rice University, Houston, Texas 77005, USA}
\affiliation{University of Virginia, Charlottesville, Virginia 22901, USA}
\affiliation{University of Washington, Seattle, Washington 98195, USA}
\author{V.M.~Abazov} \affiliation{Joint Institute for Nuclear Research, Dubna, Russia}
\author{B.~Abbott} \affiliation{University of Oklahoma, Norman, Oklahoma 73019, USA}
\author{B.S.~Acharya} \affiliation{Tata Institute of Fundamental Research, Mumbai, India}
\author{M.~Adams} \affiliation{University of Illinois at Chicago, Chicago, Illinois 60607, USA}
\author{T.~Adams} \affiliation{Florida State University, Tallahassee, Florida 32306, USA}
\author{G.D.~Alexeev} \affiliation{Joint Institute for Nuclear Research, Dubna, Russia}
\author{G.~Alkhazov} \affiliation{Petersburg Nuclear Physics Institute, St. Petersburg, Russia}
\author{A.~Alton$^{a}$} \affiliation{University of Michigan, Ann Arbor, Michigan 48109, USA}
\author{G.~Alverson} \affiliation{Northeastern University, Boston, Massachusetts 02115, USA}
\author{G.A.~Alves} \affiliation{LAFEX, Centro Brasileiro de Pesquisas F{\'\i}sicas, Rio de Janeiro, Brazil}
% ...
%
%
\collaboration{The D0 Collaboration\footnote{with visitors from
$^{a}$Augustana College, Sioux Falls, SD, USA,
$^{b}$The University of Liverpool, Liverpool, UK,
$^{c}$SLAC, Menlo Park, CA, USA,
$^{d}$ICREA/IFAE, Barcelona, Spain,
$^{e}$Centro de Investigacion en Computacion - IPN, Mexico City, Mexico,
$^{f}$ECFM, Universidad Autonoma de Sinaloa, Culiac\'an, Mexico,
and 
$^{g}$Universit{\"a}t Bern, Bern, Switzerland.%
}} \noaffiliation
\vskip 0.25cm

\date{January 4, 2011}

\begin{abstract}

We present a search for the production of a heavy gauge boson, $W^\prime$, that
decays to third-generation quarks, by the D0 Collaboration in  $p\overline{p}$ 
collisions at $\sqrt{s} = 1.96$~TeV. We set 95\% confidence
level upper limits on the production cross section times branching
fraction. For the first time, we set limits for arbitrary combinations 
of left- and right-handed couplings of the $W^\prime$ boson to fermions. For 
couplings with the same strength as the standard model $W$ boson, we set the 
following limits for $M(W^\prime) > m(\nu_R)$: $M(W^\prime)>863$~GeV for purely left-handed couplings, $M(W^\prime)>885$~GeV for purely right-handed couplings, and $M(W^\prime)>916$~GeV if both left- and  right-handed couplings are present. The limit for right-handed couplings improves  for $M(W^\prime) < m(\nu_R)$ to $M(W^\prime)>890$~GeV.

\end{abstract}

\maketitle 

In the standard model (SM), the charged weak current is mediated by 
$W$ bosons. Many models of physics beyond the standard model
predict the existence of additional charged bosons, generally called
$W^\prime$, that are more massive than the $W$ boson of the standard model.
The chiral structure of the couplings of the $W^\prime$ boson to 
fermions depends on the details of the model.  

In models of universal extra dimensions \cite{UED},  the
$W^\prime$  is the lowest Kaluza-Klein excitation of the  $W$ boson.
It therefore has the same couplings to fermions as the  $W$
boson and  couples exclusively to left-handed fermions. Left-right
symmetric models \cite{LRsym} add an additional group $SU(2)_R$ to the
SM. Thus there are two bosons, $W_L$ and $W_R$, associated
with the SM $SU(2)_L$ group and the additional $SU(2)_R$ group 
that would mix to form the physical states. One of these physical states is
identified with the  $W$ boson. Since the $W$ boson is known experimentally 
to couple only to left-handed fermions, the other state,
$W^\prime$, would have to almost exclusively couple to right-handed
fermions. If there exists a light right-handed neutrino $\nu_R$, the decay
$W^\prime\rightarrow\ell\nu_R$ (where $\ell$ is a charged lepton) 
is allowed. Otherwise, the $W^\prime$  can only
decay to quarks, making it harder to observe. 

Independent of specific models, the most general, lowest-dimension
Lagrangian for the interaction of a $W^\prime$ boson with fermion fields $f$
is given by \cite{Boos}
\begin{equation}
\lefteqn{}
{\cal L} = {V_{ij}g_w\over 2\sqrt{2}} \bar{f_i}\gamma_\mu\left(
a^R_{ij}\left(1+\gamma^5\right)+
a^L_{ij}\left(1-\gamma^5\right)\right)W^{\prime\mu} f_j  + \mbox{h.c.},
\end{equation} 
where $a^{L/R}_{ij}$ are the left/right-handed
couplings of the $W^\prime$ boson to the fermion doublet  $f_i$
and $f_j$ and $g_w$ is the weak coupling constant of the SM. If
the fermions are quarks, the $V_{ij}$ are the elements of the CKM
matrix for quark flavors $i$, $j$. If the fermions are leptons, the
$V_{ij}$ are the elements of a $3\times3$ identity matrix. 

In this Letter, we report on a search for a $W^\prime$ boson that decays
to a top quark and an anti-bottom quark 
(or charge conjugates)~\cite{Boos, Frere, Sullivan} using
2.3 fb$^{-1}$ of integrated luminosity from proton-antiproton collisions at 
$\sqrt{s}=1.96$~TeV, accumulated by the D0 experiment~\cite{run2det} at the Fermilab
Tevatron Collider between 2002 and 2007. This is the same dataset
used in the  observation of single top quark production by the D0
Collaboration \cite{singletop}. We  use the notation
$tb$ to represent the sum of the $t\bar b$ and the $\bar t {b}$ final states.
The $tb$ decay channel is sensitive to $W^\prime$ bosons with left- and 
right-handed couplings regardless of the existence of a light right-handed 
neutrino. For left-handed couplings, the process
$p\bar{p}\rightarrow W^\prime+X\rightarrow tb+X$ interferes with
SM single top quark production $p\bar{p}\rightarrow
W^*+X\rightarrow tb+X$.

We set limits on the production cross section for such $W^\prime$ bosons 
and translate them into limits on left- and right-handed 
gauge couplings. We  have previously excluded $W^\prime$ bosons with 
masses of $M({W^\prime})<731$~GeV for 
purely left-handed couplings and $M({W^\prime})<739\ (768)$~GeV for the
case of purely  right-handed couplings with (without) a light right-handed 
neutrino~\cite{Wprime_D0}. The CDF Collaboration 
has published similar limits~\cite{Wprime_CDF},
neglecting the interference with SM single top quark production. 
Relative to our earlier publication, we use more than twice the integrated 
luminosity and  improve the overall sensitivity of the analysis 
to a $W^\prime$ boson signal by using boosted decision trees (BDT).
In Ref.~\cite{singletop} 
it was demonstrated that a single top quark signal at the level 
of a few percent was clearly identifiable by the BDT technique.
For the first time, we 
perform a general analysis of the left- and right-handed gauge couplings 
$a^L_{ij}, a^R_{ij}$ of the $W^\prime$ boson, $0< a_{ij}^{L,R}<1$. 
For comparison with the earlier 
work, we also quote results for three special cases: (1) purely left-handed 
couplings, $a^L_{ij}=a, a^R_{ij}=0\ \forall i,j$; (2) purely right-handed 
couplings, $a^L_{ij}=0, a^R_{ij}=a\ \forall i,j$; and (3) an equal mixture 
of left- and right-handed couplings, $a^L_{ij}=a^R_{ij}=a\ \forall i,j$. 
Finally, we set limits on the mass of such $W^\prime$ bosons under the 
assumption of a ``SM-like'' coupling strength, i.e., $a=1$. 

We search for events in which a $W^\prime$ boson is produced and subsequently 
decays into $tb$. The top quark decays to $Wb$ and the 
$W$ boson is required to decay to $e\nu$ or $\mu\nu$.
$W$ boson decays into $\tau$-leptons or quarks suffer from an overwhelming
background. Thus our final state contains an electron or  muon, 
missing transverse momentum from the undetected neutrino, and two jets from the
fragmentation of the two $b$-quarks. Additional jets may arise from initial- or 
final-state radiation. We acquire these 
events using a logical OR of many trigger conditions 
that require a combination of jets, electrons and muons. 

Object selections and definitions are identical to Ref.~\cite{singletop}.  
Events are selected with exactly one isolated electron with $p_T > 15$~GeV 
for two-jet events or $p_T > 20$~GeV for three- and four-jet events, 
and $|\eta| < 1.1$ or one isolated muon with $p_T > 15$~GeV and $|\eta| < 2.0$, 
where where $\eta$ is the pseudorapidity.
The leading jet must have $p_T>25$~GeV and there must be a total of two, 
three, or four jets with $p_T>15$~GeV and $|\eta|<3.4$.
Jets are defined using a cone 
algorithm~\cite{cone-algorithm}  with a cone radius 
$\Delta \mathcal{R}=\sqrt{(\Delta\phi)^2+(\Delta y)^2}=0.5$, 
where $y$ is the rapidity. 
Missing $p_T$ is required to be greater than 
$20$~($25$)~GeV for two-jet (three- or four-jet) events.
To enhance the signal content of the selected data sample, one or two of the
jets are required to be identified as originating from the fragmentation of a 
$b$ quark~\cite{btagging}. We call these jets $b$-tagged.
In order to minimize the effect of  interference with SM single top quark production, we require the invariant mass, 
$\sqrt{ \hat{s} }$, of the reconstructed charged lepton, the jets, and 
missing $p_T$ to be greater than 400~GeV.

We simulate the complete chain of $W^\prime$ decays, taking into account 
finite widths and spin correlations between the production of resonance 
states and their decay using the {\sc singletop} Monte Carlo 
package~\cite{Boos:2006af} based on the  {\singletop} event 
generator~\cite{Boos:2004kh}. We set the top quark mass to 172.5~GeV; 
the factorization scale is set to $M({W{^\prime}})$ and we use the 
CTEQ6M  parton distribution functions~\cite{CTEQ6M}.
The {\singletop} generator reproduces the next-to-leading-order (NLO) 
kinematic distributions,  and the interference term between the $W{^\prime}$ boson signal 
and the SM $W$~boson. The width of the $W{^\prime}$ boson varies 
between 20 and 30~GeV for $W{^\prime}$~masses between 600 and 
900~GeV~\cite{Boos, Sullivan}. It is 25\% smaller without a light 
right-handed neutrino. The width does not have a significant effect 
on our search because it is smaller than the detector resolution. 
Therefore, the only relevant effect
of a massive neutrino is the larger branching fraction for 
$W{^\prime}\rightarrow tb$. 
Unless mentioned otherwise we will assume in the following that there is a
right-handed neutrino with $M(W')>m(\nu_R)$. In this case the limits for 
right-handed couplings will be  weaker than for $M(W')<m(\nu_R)$ 
because of the smaller branching fraction for $W'\rightarrow tb$.
In the absence of interference between $W$
and $W{^\prime}$~bosons and in the presence of  a light right-handed neutrino, 
there is no difference between $W^\prime$ bosons with left- and right-handed 
couplings for our search. To probe for $W{^\prime}$ bosons with masses above 
the currently published limits, we simulate $W^\prime$ bosons at nine mass 
values from 600 to 1000~GeV.

Background yields are estimated using both Monte Carlo samples and data. 
The procedure is identical to that used in Ref.~\cite{singletop}. 
The multijet background arises from events in which a jet 
mimics the signature of an isolated charged lepton. 
It is modeled using control 
data samples with non isolated leptons. 
The dominant background for our search originates from $W$+jets production. 
Smaller sources of backgrounds are from  $t\overline t$ pair, $t$-channel 
single top  quark ($tqb$), diboson, and $Z$+jets events. 
The diboson ($WW$, $WZ$, $ZZ$) backgrounds  
are modeled using {\sc pythia}~\cite{pythia}. The other background processes 
are modeled using the {\sc alpgen}~\cite{alpgen} event generator 
and subsequently hadronized using {\sc pythia}. 
The fraction of $W$+heavy flavor events ($Wb\bar{b}$, and $Wc\bar{c}$)
is obtained by scaling the cross sections calculated by {\sc alpgen} 
to the NLO cross section by 
a factor of 1.47. An additional  correction factor of $0.95 \pm 0.13$, 
derived from data samples  with different number of $b$-tagged jets, 
is applied. 
The $Wc+$light parton cross sections are scaled by a factor of 1.38.
All processes except the $W$+jets and multijets
background are normalized to their expected cross sections~\cite{bkgxsec,singletop_xsec}.
The $W$+jets and multijets yields are obtained by normalizing 
to the data sample after 
subtracting all other backgrounds and before selecting events based on 
$b$-tagged jets. At this point in the selection, 
the data are completely dominated by background and any contamination from 
a $W^\prime$ boson signal is negligible.

The sensitivity of the search is maximized by dividing the data into 24 
independent channels based on lepton flavor ($e$, $\mu$), jet multiplicity 
(2, 3, 4), number of $b$-tagged jets (1, 2), and two data collection 
periods, 
to take into account different signal acceptances and signal-to-background 
ratios~\cite{singletop}. After applying all selections, we find the event yields for data and 
backgrounds as shown in Table~\ref{tab:event-yields}. 
The requirement $\sqrt{\hat{s}}>400$~GeV accepts most of the $W^\prime$
contribution but eliminates most of the $W$ boson contribution. 
As the mass of the $W^\prime$ boson increases, its contribution to the 
$tb$ final state decreases relative to that of the $W$ boson, 
and therefore, the efficiency of this requirement decreases 
between 3.5\% and 0.5\% for $W^\prime$ boson masses between 600 and 1000~GeV.

\begin{table}[!h!tbp]
\begin{center}
\caption{Data and SM background  event yields with systematic uncertainties.}\label{tab:event-yields}
\begin{tabular}
{l  r@{ $\pm$ }l }
\hline\hline
Process &\multicolumn{2}{c}{ Events }\\\hline
$tqb$               & {$26.4$} & {2.5} \\
$t\bar{t}$          & {$424.7$}& {58.4} \\
$W$+jets            & {$279.5$}& {18.3} \\
$Z$+jets            & {$26.0$} & {3.2}\\
Dibosons             & {$13.0$} & {1.6}\\
Multijets           & {$60.5$} & {10.8} \\ \hline
Total background    & {$830$}& {62}\\
\hline
Data                &  \multicolumn{2}{c}{$831$}   \\ \hline\hline
\end{tabular}
\end{center}
\end{table}

\begin{figure*}[ht]
\begin{center}
{\includegraphics[scale=0.28]{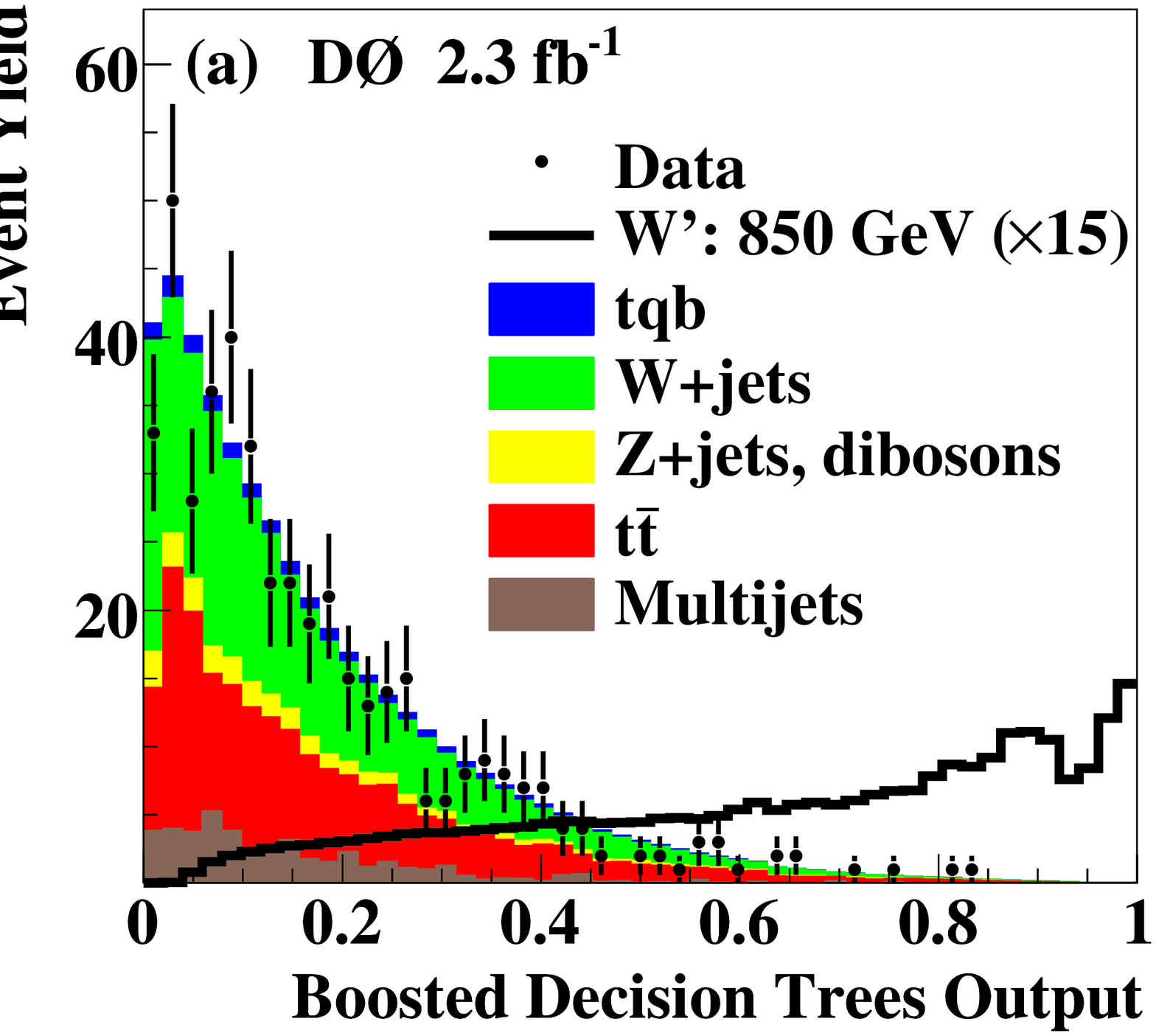}}
{\includegraphics[scale=0.28]{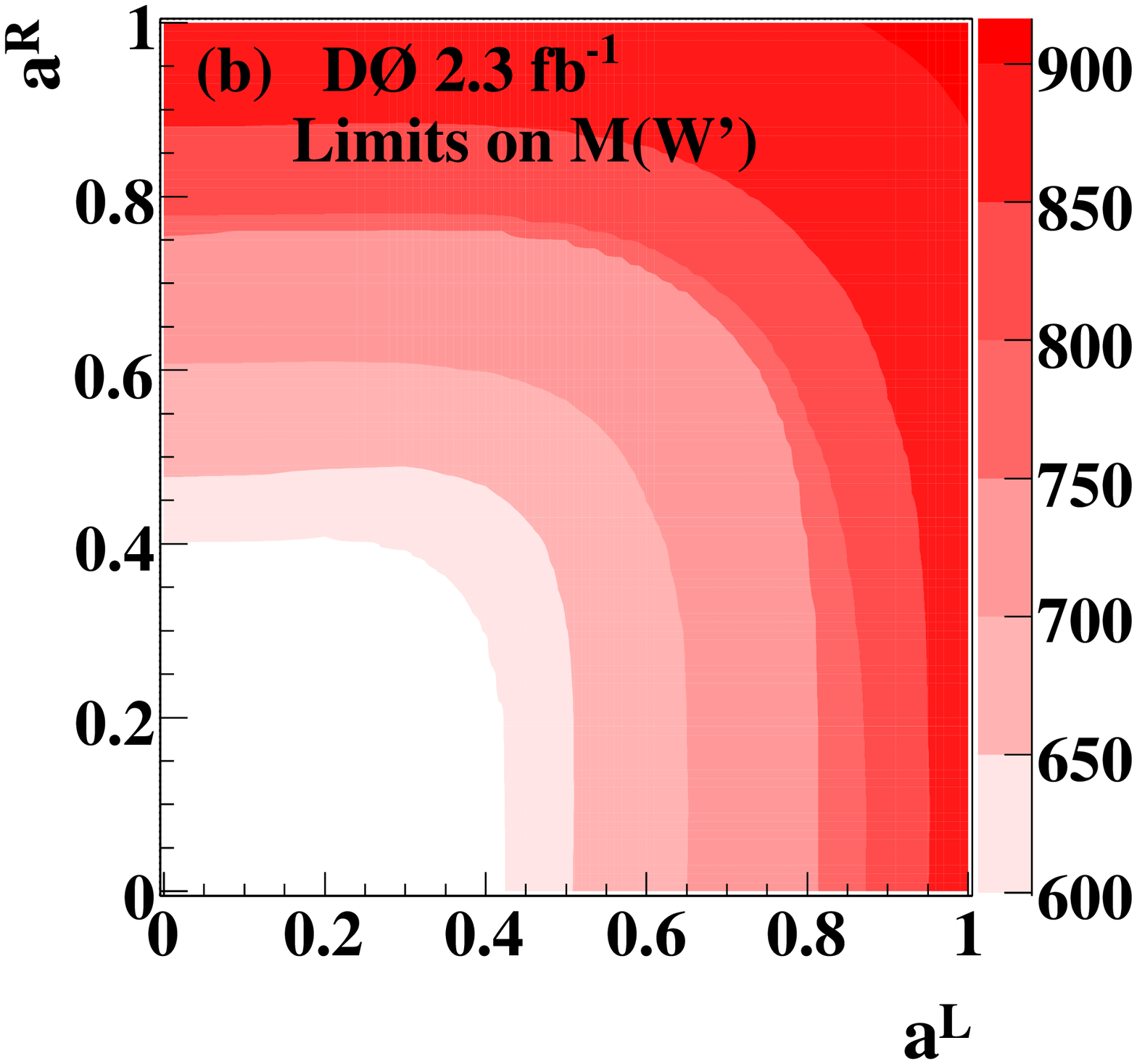}}
{\includegraphics[scale=0.28]{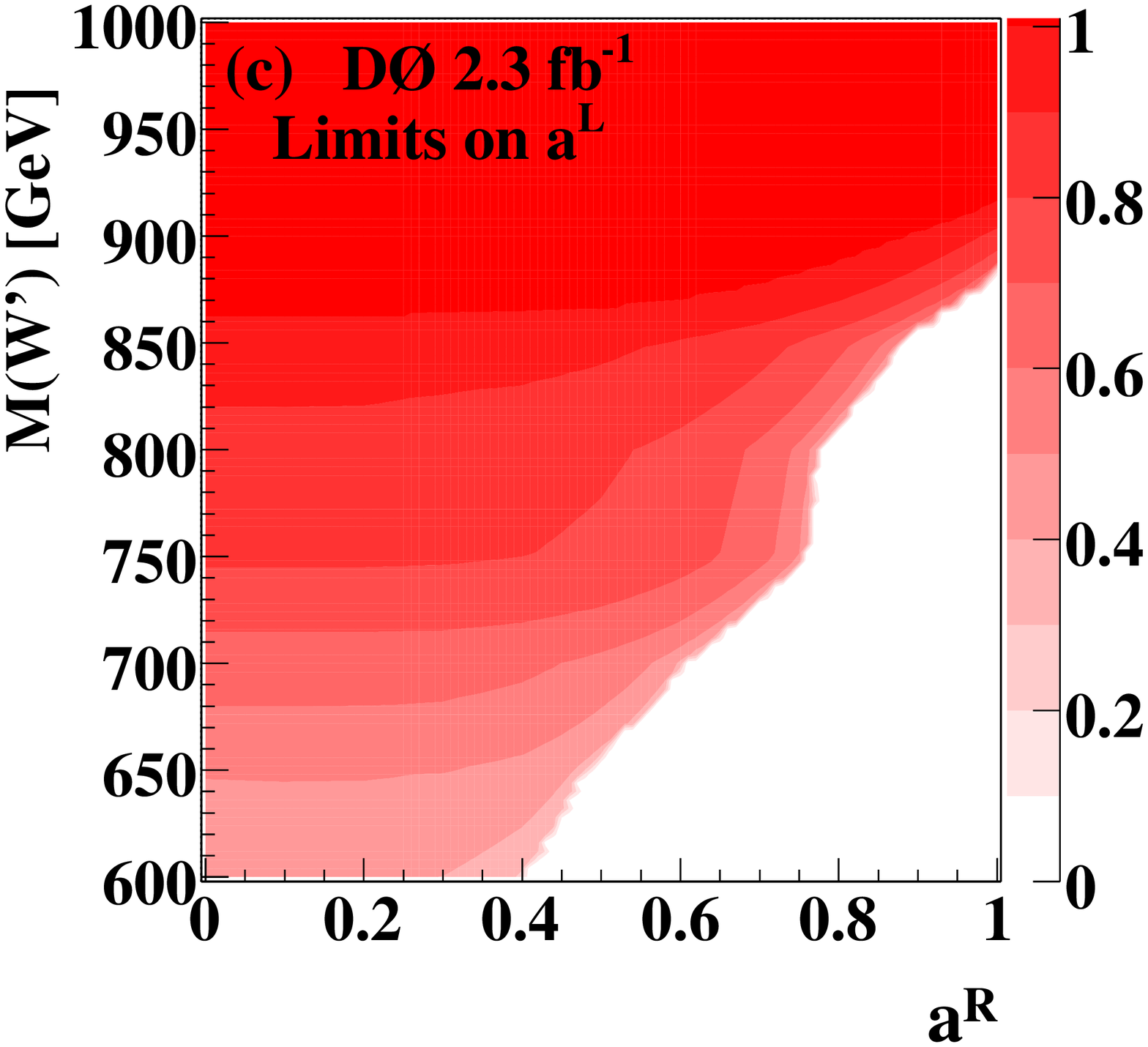}}
\caption{(a) Distribution of the discriminant for data (points) compared to 
the background model summed over all analysis channels (filled histograms) 
and expected $W^\prime$ boson signal with mixed couplings $(a^L,a^R)=(1,1)$ (open histogram),
(b) Contour plots of 95\% C.L. lower limits on $M(W^\prime)$ 
in the ($a^L,a^R$) plane, and  (c) 95\% C.L. upper limits on the coupling $a^L$ 
in the ($a^R,M(W^\prime)$) plane. Here $M(W^\prime) > m(\nu_R)$.
\label{fig:disc_limit}}
\end{center}
\end{figure*}

At this stage, the expected $W^\prime$ boson signal would constitute only a 
small 
fraction of the selected data sample. To improve discrimination, we calculate a
multivariate discriminant based on BDTs that separates
the $W^\prime$ boson signal from background
and thus enhances the probability to observe $W^\prime$ boson production. 
We compute the discriminant 
independently in each of the 24 channels. The input variables used to train 
the BDT discriminants take into account the kinematic properties and angular 
correlations of the reconstructed objects and the topology of the event. 
A BDT is trained for each $W^\prime$ boson mass value using the Monte Carlo 
sample generated with $a^L=a^R=1$. The final discriminant for all 24 channels 
combined is shown in Fig.~\ref{fig:disc_limit}(a). 
To represent the discriminant distribution expected from an arbitrary 
combination of couplings, we combine samples of $W^\prime$ decays generated 
with left-handed, right-handed and mixed couplings and SM $s$-channel $tb$ 
production based on theoretical expectations from Ref~\cite{Boos}. 
Background events  
preferentially populate the low discriminant region whereas signal events are 
clustered towards high discriminant values. We observe good agreement between 
background prediction and data in all channels. The data show no deviation from 
background only expectations  
and we set 95\% C.L. upper limits on the cross section for the process 
$W^\prime \rightarrow t b \rightarrow \ell\nu bb$ using 
Bayesian statistics~\cite{top-stat}.  
A Poisson distribution for the observed counts and
a flat non-negative prior probability for the signal cross section are assumed. 
Systematic uncertainties are taken into account with Gaussian priors.

The systematic uncertainties that affect the signal and background
models are  described in Ref.~\cite{syst}. The largest sources of
systematic uncertainties are the jet energy scale calibration and the
modeling of $b$-tagging performance. Smaller uncertainties arise from
the finite size of the MC samples, the corrections of the flavor
composition of $W$+jets events, and from the normalization of the
background. The total uncertainty in the background  yield varies
between 8\% and 10\% for the different channels. In determining the
effect of the uncertainties from jet energy scale calibration, $b$-tag
modeling, and $W$+jets modeling, we take into account changes in the
shape of the discriminant distributions in addition to normalization
effects. 

The cross section for single top quark production in the presence of a $W'$
boson for any set of coupling 
values can be written in terms of the cross sections ${\sigma}_L$ for purely 
left-handed  couplings $(a^L,a^R)=(1,0)$, ${\sigma}_R$ for purely 
right-handed couplings $(a^L,a^R)=(0,1)$, ${\sigma}_{LR}$ for mixed 
couplings $(a^L,a^R)=(1,1)$, and  ${\sigma}_{SM}$ for SM couplings 
$(a^L,a^R)=(0,0)$. It is given by:
\begin{eqnarray}\label{eq:xsec}
{\sigma} &=& {\sigma}_{SM} + a^L_{ud}a^L_{tb}
\left({\sigma}_L - {\sigma}_R \right) \\ \nonumber
         &+& \left(\left(a^L_{ud} a^L_{tb}\right)^2 
          +        \left(a^R_{ud} a^R_{tb}\right)^2\right)
\left({\sigma}_R- {\sigma}_{SM}\right) \\ 
         &+& \frac{1}{2}\left(\left(a^L_{ud} a^R_{tb}\right)^2 
          +                   \left(a^R_{ud} a^L_{tb}\right)^2\right)
\left( {\sigma}_{LR} - {\sigma}_L - {\sigma}_R  + {\sigma}_{SM} \right). 
\nonumber
\end{eqnarray}
The predicted cross section for SM single top quark production through $s$-channel $W$ boson exchange is 
$\sigma_{SM}=1.12\pm0.05$~pb~\cite{singletop_xsec}. 
The predicted cross sections ${\sigma}_L$, 
${\sigma}_R$, and ${\sigma}_{LR}$ as a function of $W^\prime$ boson mass are 
listed in Table~\ref{tab:xs}. We assume that the couplings to first 
generation quarks, $a_{ud}$, which are important for the production of the 
$W^\prime$ boson, and the couplings to third generation quarks, $a_{tb}$, 
which are important for the decay of the $W^\prime$ boson, are equal. 
For given values of $a^L$ and $a^R$, the distributions are obtained by 
combining the four samples according to Eq.~(\ref{eq:xsec}). 

\begin{figure*}[ht]
\begin{center}
{\includegraphics[scale=0.29]{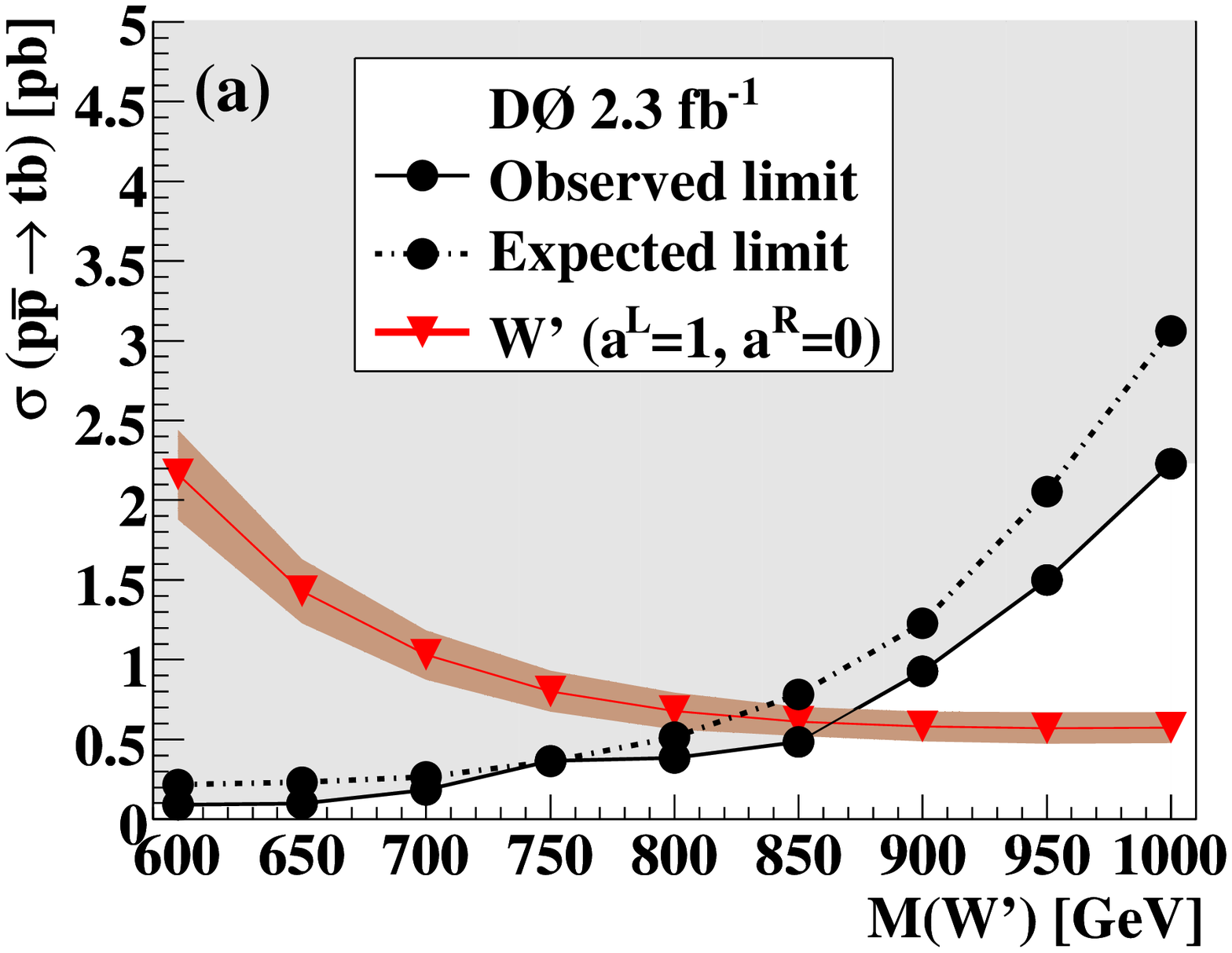}}
{\includegraphics[scale=0.29]{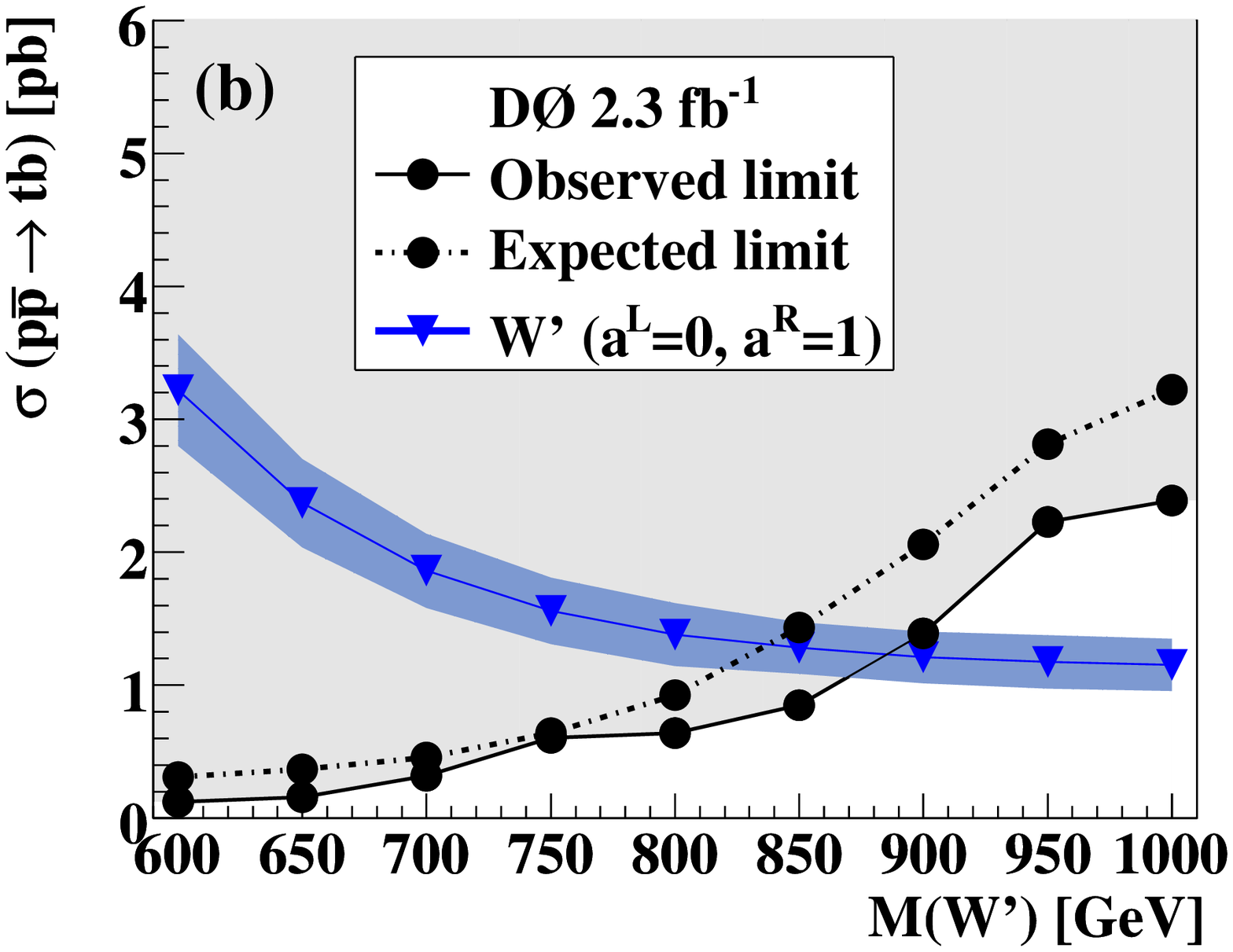}}
{\includegraphics[scale=0.29]{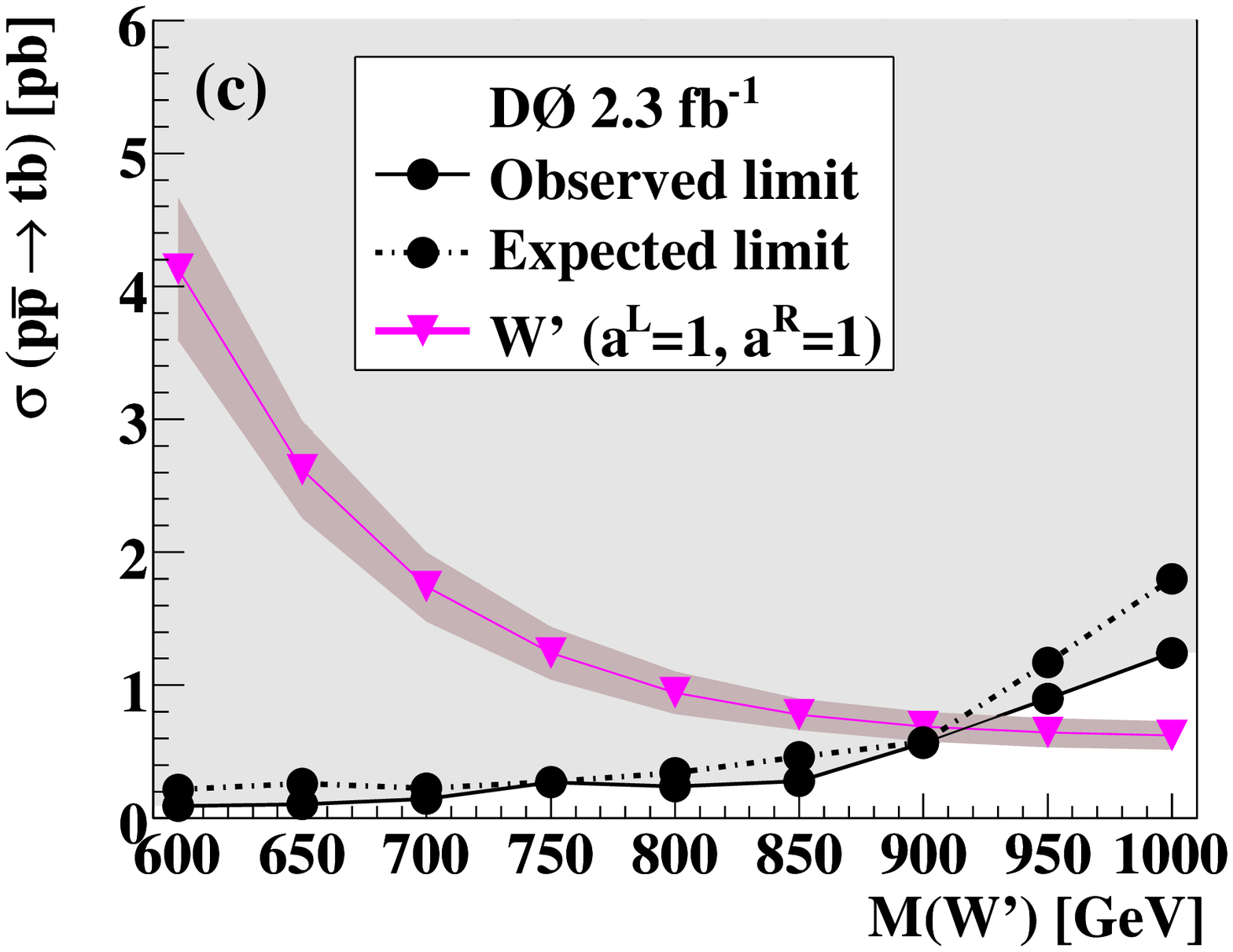}}
\caption{Expected and observed upper limits and 
theory prediction for the production cross section (the shaded band indicates
the uncertainty~\cite{Sullivan}) for 
$W^\prime$ bosons with (a) left-handed couplings $(a^L,a^R)=(1,0)$, (b) right-handed couplings $(a^L,a^R)=(0,1)$ and, (c) mixed  couplings $(a^L,a^R)=(1,1)$  as a function of the $W^\prime$ boson mass.  Here $M(W^\prime) > m(\nu_R)$.
\label{fig:limitsLRM}}
\end{center}
\end{figure*}

We vary both $a^L$ and $a^R$ between 0 and 1 in steps of 0.1, for each 
$W^\prime$ boson mass value. We consider each of these 121
combinations of $a^L$, $a^R$,  and $M(W^\prime)$ as a model.
For each of these models, the ratio of the $s$-channel and 
the $W^\prime$ cross sections is fixed. We then vary the total cross section to 
determine the expected and observed 95\% C.L.  upper limits.
For the three special cases $(a^L,a^R)=(1,0)$, $(0,1)$, and $(1,1)$,
the measurements are listed along with the theoretical cross sections 
in Table~\ref{tab:xs}.  For small $W^\prime$ masses, the cross section
limits drop below the cross section for SM single top quark production 
in the $s$-channel, because the data do not accommodate the large $W^\prime$ 
component that would have to be present for these models. 
The cross section limit reflects the maximum total cross section that the 
data can accommodate for the ratio of $s$-channel and $W^\prime$ production 
in the particular model.
We can now assume values for any two of the
three  parameters, $a^L$, $a^R$, and $M(W^\prime)$, and interpolate the
cross section limit in  the third parameter value. The value of the
third parameter at which  the cross section limit equals the theory
cross section~\cite{Boos}  represents the limit on the third
parameter.  
Figure~\ref{fig:disc_limit}(b) shows  contours of lower limits 
for the $W^\prime$ boson mass  in the
($a^L,a^R$) plane, and Fig.~\ref{fig:disc_limit}(c) shows 
contours of upper limits for the coupling $a^L$ in the 
($a^R,M(W^\prime)$) plane for $M(W^\prime) > m(\nu_R)$.
Fig.~\ref{fig:limitsLRM} shows the cross section limits for the 
three special cases.
As the $W^\prime$ boson mass increases, the 
selection efficiency 
decreases and the upper limits on the cross section increase.

\begin{table}
\caption{\label{tab:xs} 
NLO production cross section times branching fraction, $\sigma(p\bar{p}\rightarrow W/W^\prime\rightarrow tb)$, and expected and observed 95\% C.L. upper limits for different $W^\prime$ boson masses, assuming $M(W^\prime) > m(\nu_R)$, in pb.}
\begin{center}
\begin{tabular}{lccccccccc} \hline \hline
$M(W{^\prime})$ & \multicolumn{3}{c}{$(a^L,a^R)=(0,1) $} &  \multicolumn{3}{c} {$(a^L,a^R)=(1,0)$} &  \multicolumn{3}{c} {$(a^L,a^R)=(1,1)$} \\
(GeV)& ${\sigma}_R$ &  obs & exp  & ${\sigma}_L$ & obs & exp & ${\sigma}_{LR}$ & obs & exp\\ \hline
~600  &  3.22 &  0.12 & 0.31  & 2.16 & 0.09 & 0.22& 4.13  & 0.09 & 0.21   \\
~650  &  2.37 &  0.16 & 0.37  & 1.43 & 0.10 & 0.23& 2.62  & 0.10 & 0.26   \\
~700  &  1.86 &  0.32 & 0.46  & 1.03 & 0.18 & 0.26& 1.74  & 0.15 & 0.23   \\
~750  &  1.56 &  0.60 & 0.64  & 0.80 & 0.37 & 0.36& 1.24  & 0.27 & 0.28   \\
~800  &  1.38 &  0.64 & 0.92  & 0.68 & 0.39 & 0.51& 0.95  & 0.24 & 0.34   \\
~850  &  1.28 &  0.85 & 1.44  & 0.61 & 0.48 & 0.78& 0.78  & 0.28 & 0.46   \\
~900  &  1.21 &  1.39 & 2.06  & 0.58 & 0.93 & 1.23& 0.69  & 0.56 & 0.57   \\
~950  &  1.18 &  2.23 & 2.81  & 0.57 & 1.50 & 2.05& 0.64  & 0.90 & 1.17   \\
1000 &   1.15 &  2.39 & 3.22  & 0.57 & 2.23 & 3.06& 0.62  & 1.24 & 1.80   \\
\hline \hline
\end{tabular}
\end{center}
\end{table}

In conclusion, we have carried out a search for a massive charged gauge boson, 
$W^\prime$, that decays to $tb$. We considered a model-independent 
approach in which the $W^\prime$ boson may couple to fermions with any combination
of left- and right-handed couplings. We do not observe any  
deviations from 
SM expectations and set upper limits on the cross section for the 
production of $W^\prime$ bosons. We compare upper limits 
to theory cross sections to 
extract the 
following limits for $M(W^\prime) > m(\nu_R)$: $M(W^\prime)>863$~GeV for purely left-handed couplings, $M(W^\prime)>885$~GeV for purely right-handed couplings, and $M(W^\prime)>916$~GeV if both left- and  right-handed couplings are present. The limit for right-handed couplings improves for $M(W^\prime) < m(\nu_R)$ to $M(W^\prime)>890$~GeV.
These mass limits improve previously  published results by more than 100~GeV.

%
We thank the staffs at Fermilab and collaborating institutions,
and acknowledge support from the
DOE and NSF (USA);
CEA and CNRS/IN2P3 (France);
FASI, Rosatom and RFBR (Russia);
CNPq, FAPERJ, FAPESP and FUNDUNESP (Brazil);
DAE and DST (India);
Colciencias (Colombia);
CONACyT (Mexico);
KRF and KOSEF (Korea);
CONICET and UBACyT (Argentina);
FOM (The Netherlands);
STFC and the Royal Society (United Kingdom);
MSMT and GACR (Czech Republic);
CRC Program and NSERC (Canada);
BMBF and DFG (Germany);
SFI (Ireland);
The Swedish Research Council (Sweden);
and
CAS and CNSF (China).

\end{document}